\documentclass[twocolumn,showpacs,preprintnumbers,amsmath,amssymb]{revtex4}
\usepackage{dcolumn}
\usepackage{bm}
\usepackage{epsfig}
\begin{document}
\title{Controlled Split-Recombination of 2D Matter-Wave Solitons in Time-Dependent Trap}
\author{V. Ramesh Kumar$^1$}
\author{Lin Wen$^1$}
\author{R. Radha$^2$}
\author{W. M. Liu$^1$}
  \affiliation{$^1$Beijing National Laboratory for Condensed Matter Physics, Institute of Physics, Chinese Academy of Sciences, Beijing, China\\
$^2$Centre for Nonlinear Science, Govt. College for Women
(Autonomous), Kumbakonam, India}

\begin{abstract}We propose a novel approach to manipulate two-dimensional bright matter-wave solitons
by tuning the frequency of the trap which is different from
Feshbach resonance technique. The exact bright soliton solutions
for two-dimensional Gross-Pitaevskii (GP) equation with attractive
interaction strength in a time-dependent trap are constructed
analytically and its dynamics show no collapse while modulating
the trap frequency. The two-soliton dynamics exhibits an
interesting splitting and recombination phenomenon which generates
interference pattern in the process. This type of behaviour in
two-dimensional BECs has wider ramifications and our approach
opens new avenues in stabilizing bright solitons in higher
dimensional regime. We have also explored the experimental
realization of this novel phenomenon.

\end{abstract}


\maketitle

\section{Introduction}
A Bose-Einstein condensate (BEC) is a state of matter in which a
large number of bosonic atoms occupy the lowest quantum state of
the external potential at extremely low temperatures near absolute
zero, allowing quantum effects to be observed on a macroscopic
scale. The experimental realization of BECs in weakly interacting
gases has opened the floodgates in the field of atom optics and
condensed matter physics. The collective excitation of
matter-waves in BECs like matter-wave solitons
\cite{burge,densc,strec,gubes}, periodic waves \cite{abdull},
shock waves \cite{perez},  vortices \cite{ander01} and necklaces
\cite{theoc} has generated a lot of interest on the exploration of
the dynamics of BECs from both experimental and theoretical
perspectives. Even though the experimental realization of
Bose-Einstein condensation in different atomic systems has been
observed, the stability of BECs imposes restriction on the
observation of dynamical properties of such topological
excitations. This is because of the higher dimensionality of the
system. For example, attractive Li BECs have been shown to
collapse in three dimensions \cite{bradl2,sacke}. But, the
one-dimensional attractive Li BECs in a standing light wave
potential are stable \cite{brons}. This is primarily because of
the fact that in higher dimensional BECs, a nonlinear excitation
of atoms just bigger than the background is susceptible to even
smaller perturbations. The dark solitons will decay into vortex
pairs under small transverse perturbation \cite{ander01}.
Therefore, longtime dynamical behaviour of nonlinear excitations
in higher dimensional BECs is hard to observe in real experiments.

However, it was identified that the bright solitons and vortex
solitons can be stabilized without trap in higher dimensional BECs
by rapidly oscillating atom-atom interaction strength using
Feshbach resonance technique \cite{saito,abdullaev,adhik} where
resonantly tuning uniform magnetic field produces an effective
confinement creating stable self-confined condensates. Similarly,
a stable shallow ring dark soliton has also been observed in
two-dimensional BECs with tunable interaction \cite{hu09}. Thus,
to stabilize the solitons in BECs, one has to change the
interaction strength between the atoms by tuning magnetic field
near Feshbach resonance \cite{cornish,inouye}.

Can one suitably modulate the trap frequency in a time-dependent
trap and control the stability of the condensates instead of
varying the interaction strength? This paper is aimed at the
investigation of the (2+1) dimensional GP equation with
time-dependent trapping potential describing the dynamics of
matter waves in pancake-shaped BECs. The fact that the trap
frequency varies with time means that one can suitably tune the
time-dependent harmonic trap to stabilize the bright solitons and
this is an alternative way of stabilizing the bright solitons in
BECs rather than tuning the interaction strength. We first
construct exact bright matter-wave soliton solutions using Hirota
method \cite{hirot} and then study the impact of modulating the
trap frequency. We demonstrate split-recombination phenomenon by
modulating the trap frequency. The frequency modulation technique
is well known in controlling molecular BEC scattering length and
condensate spilling \cite{jochim}. Our results open a new window
to manipulate atoms in higher dimensional BECs.

\section{Model}

At ultra low temperatures, the dynamics of BECs can be described
by the three dimensional time-dependent GP equation of the
following form \cite{dalfo}
\begin{equation}
i\hbar \frac{\partial \Psi(\mathbf{r},t)}{\partial t}=\left(-\frac{\hbar^2}{%
2m}\nabla^2 +V(\mathbf{r},t)+U|\Psi(\mathbf{r},t)|^2 \right)\Psi(\mathbf{r}%
,t)
\end{equation}
where $\Psi[\textbf{r}(x,y,t),t]$ represents the
condensate wave function normalized by the particle number $N = \int d%
\mathbf{r} |\Psi|^2$, $\nabla^2$ denotes the Laplacian operator and $V(\mathbf{r%
})$ is the trapping potential of the form $V(\mathbf{r}) =
m(\omega_r(t)^2 r^2 + \omega_z^2 z^2)$, where $r^2 = x^2+y^2$,
$\omega_{r,z}$ are the confinement frequencies in the radial and
axial directions, respectively. The coefficient $U$ is related to
s-wave scattering length $a_s$ controlled by Feshbach resonance as
$U = 4 \pi \hbar^2 a_s/m$, $m$ is the atom mass.

The three-dimensional time-dependent GP equation effectively
reduces to two-dimensional GP equation when the radial trapping
frequency $\omega_r$ is very small compared to axial trapping
frequency $\omega_z$, i.e ($|\omega_r|/\omega_z\ll 1$).
Consequently, the atoms move freely in the $x$, $y$ plane.
Allowing the radial trapping frequency to vary with time i.e.
$\omega_r = \omega_r(t)$, the corresponding two-dimensional GP
equation becomes
\begin{equation}
i\frac{\partial\psi}{\partial
t}+\frac{1}{2}\left(\frac{\partial^2}{\partial
x^2}+\frac{\partial^2}{\partial y^2}\right)\psi+g |\psi|^2 \psi +
\Omega(t)^2 r^2\psi=0,
\end{equation}
where the strength of trapping potential is related to radial and
axial trapping frequency as $\Omega(t)^2 =
|\omega_r(t)|^2/\omega_z^2 $ and $g = (8\pi m \omega_z
/\hbar)^{1/2}N a_s$ represents the strength of attractive
interaction ($g
> 0$). In Eq. (2) the length,
time, frequency and wave function are measured in units of $a_h$,
$1/\omega_{r0}$, $\omega_{r0}$, and $N^{1/2}/{a_h}$, respectively,
where $a_h = \sqrt{\hbar/m\omega_{r0}}$ and $\omega_{r0}$ is a
radial trapping frequency at $t=t_0$.  When $\Omega(t)^2 > 0$, the
trapping potential is expulsive and $\Omega(t)^2 < 0$ represents
the confining potential.

The nature of the nonlinear excitations and stability of pan cake
shaped BECs in unmodulated traps (time independent traps) is well
studied in different physical contexts \cite{abdulll,luca,law}.
However, the impact of modulating the trap frequency on the
condensates is not yet known. So, the investigation of (2+1) GP
equation in an arbitrary time dependent trapping potential with
constant attractive interaction strength assumes tremendous
significance.

\section{One soliton solution}
The aim of this paper is to solve the (2+1) GP equation for an
arbitrary time-dependent trapping potential with constant
attractive interaction strength. To solve Eq. (2), we use Hirota
bilinear method. To generate Hirota bilinear form of Eq. (2), we
make use of the following transformation of the form
\begin{equation}
\psi = e^{i \alpha(t) \frac{(x^2 +y^2)}{2}}\frac{G}{F},
\end{equation}
where $G = G(x,y,t)$ is complex, while $F = F(x,y,t)$ is real.
Substituting Eq. (3) into Eq. (2), one obtains the following
Hirota bilinear form,
\begin{eqnarray}
(i D_t + \frac{1}{2} (D_x^2 + D_y^2) &+& i \alpha(t) x D_x +
i\alpha(t) y D_y
\nonumber \\
&+& i \alpha(t))G\cdot F = 0, \\
\frac{1}{2}(D_x^2 &+& D_y^2) F\cdot F = g |G|^2,
\end{eqnarray}
and the trapping potential $\Omega(t)^2$ is related to the
function $\alpha(t)$ as
\begin{equation}
2 \Omega(t)^2 = \frac{d\alpha(t)}{dt}+\alpha(t)^2.
\end{equation}
It should be mentioned that the choice of trapping potential
strength depends on the solvability of Ricatti equation (Eq. (6)),
but the solution of Ricatti equation is not unique. One could also
get different physically realizable potential by properly choosing
the time dependent function $\alpha(t)$.

The operator $D_x$ and $D_t$ are defined as
\begin{eqnarray}
D_t^m D_x^m (G\cdot F) = &&\left(\frac{\partial}{\partial t}-\frac{\partial}{%
\partial t^{\prime }}\right)^m \left(\frac{\partial}{\partial x}-\frac{%
\partial}{\partial x^{\prime }}\right)^n \times  \nonumber \\
&& G(t,x) F(t^{\prime },x^{\prime })|_{t^{\prime }=t, x^{\prime }=
x}.
\end{eqnarray}
The functions $G$ and $F$ can be extended as
\begin{eqnarray}
G= \varepsilon \bar{g}^{(1)} + \varepsilon^3 \bar{g}^{(3)} +
\varepsilon^5
\bar{g}^{(5)}+\cdots, \\
F = 1+\varepsilon^2 f^{(2)} + \varepsilon^4 f^{(4)}+\cdots.
\end{eqnarray}
Substituting $G$ and $F$ into the bilinear forms and collecting
the coefficient of various powers of $\varepsilon$, one obtains
the system of linear partial differential equations which can be
recursively solved. The function $\bar{g}^{(1)}$ can be
represented as the plane wave solution of the form
\begin{equation}
\bar{g}^{(1)}= \sum_{j=1}^{N} e^{\chi_j}, \quad\quad \chi_j =
h_1^{(j)}(t) x + h_2^{(j)} (t) y + h_3^{(j)}(t).
\end{equation}

To generate one soliton solution, we take $N = 1$  to obtain $\bar{g}^{(1)}$ and $%
f^{(2)} $ as

\begin{eqnarray}
\bar{g}^{(1)} &&= e^{\chi_1(x,y,t)}, \\
f^{(2)} &&= e^{\chi_1(x,y,t) + \chi_1^*(x,y,t) + \eta_1(t)},
\end{eqnarray}
with
\begin{eqnarray}
\chi_1(x,y,t) &&= h_1^{(1)}(t) x +h_2^{(1)}(t) y+h_3^{(1)}(t), \\
h_1^{(1)}(t) &&= a e^{-\int\alpha(t)dt}, \quad h_2^{(1)}(t) = b
e^{-\int\alpha(t)dt},  \nonumber \\
h_3^{(1)}(t)&& = \int(\frac{i}{2} (a^2+b^2)
e^{-2\int\alpha(t)dt}-\alpha(t))dt, \\
e^{\eta_1(t)}&&= g\frac{ e^{2 \int
\alpha(t)dt}}{(a+a^*)^2+(b+b^*)^2},
\end{eqnarray}
where $a$ and $b$ are complex constants, and $\chi_1^*$ denotes
complex conjugate of $\chi_1$, $\alpha(t)$ is real function. Substituting $\bar{g}^{(1)}$ and $%
f^{(2)}$ into the truncated series, we obtain $G = \bar{g}^{(1)}$
and $F = 1+f^{(2)}$. Thus, we get the one soliton solution of the
following form

\begin{equation}
\psi_1 = \frac{1}{2}e^{-\eta_1/2}sech\left(Re(\chi_1)+\frac{\eta_1}{2}%
\right)e^{i Im(\chi_1)+i \alpha(t)/2 (x^2+y^2)}.
\end{equation}

From the above solution, we observe that the amplitude $(e^{-2\int
\alpha(t) dt}(a_R^2+b_R^2)/4g)^{1/2} $,
velocity(Re($h_3(t))+\eta_1(t)/2$),
position(Re($\chi_1)+\eta_1/2$) and phase(arg($\psi$)) of the
soliton depend on the frequency of the trapping potential (i.e.
$2\omega_r(t)^2/\omega_z^2=2\Omega(t)^2 =\alpha'(t)+\alpha(t)^2$)
and attractive interaction strength $g$. Since the frequency of
the trapping potential could be modulated with respect to time,
one can easily control the physical properties of bright solitons
and hence the condensates. We can choose the trapping potential to
be either confining ($\Omega(t)^2<0$) or expulsive
($\Omega(t)^2>0$). And, we can also continuously change the nature
of the trapping potential from confining to expulsive or expulsive
to confining. It is interesting to note that this exact bright
soliton solution involves constant attractive interaction strength
which is independent of trapping potential strength unlike the
quasi one-dimensional BECs \cite{mypra,serkin,liu}, wherein one
needs a delicate balance between time-dependent trap and
interaction strength.

The time-dependent trapping frequency allows us to choose
different physically realizable potentials and study the effect of
trap on the dynamics of solitons. For example, in the case of a
constant trapping potential ($\Omega(t)^2=\Omega_0=0.4$), the
amplitude and width of the soliton varies with time as shown in
Fig. 1(a), in which the amplitude of the soliton increases and
reach the maximum value which depends on the strength of trapping
potential and then decreases. When the trapping potential is
suddenly switched off, the amplitude decreases and the width
increases as shown in Fig. 1(b). From these two cases, we observe
that the time-independent trap and zero-trapping potential cannot
keep the soliton as stable. As an alternative, one can look to
stabilize the bright solitons by modulating the trapping frequency
with time. For example, choosing $\Omega(t)^2=-2-\cos(2t)$, the
$\alpha(t)$ takes the following form
\begin{equation}
\alpha(t)=\frac{ mathieuC'(4,-1,t)+mathieuS'(4,-1,t)}{
mathieuC(4,-1,t)+mathieuS(4,-1,t)}.\nonumber
\end{equation}
The mathieuC($l,m,t$) and mathieuS($l,m,t$) are even and odd
functions, respectively. When $m =0$, they are simply as
$\cos(\sqrt{l}t)$ and $\sin(\sqrt{l}t)$ where the prime functions
are $t$ derivatives of matheiu functions. When we modulate the
trap strength periodically by tuning the radial trapping frequency
($\omega_r (t)$) within the two-dimensional confining regime, the
amplitude and width of the soliton oscillates and gets amplified
as shown in Fig. 1(c), but does not collapse or dilute even if one
waits long enough. In this case, the stability of the soliton is
partial. One can choose the other possible time-dependent
potentials to stabilize the soliton where the amplitude and width
of the soliton remains constant or oscillates without
amplification.

When the confining trapping potential strength decreases
exponentially ($\Omega(t)^2=e^{-4 t} (0.1352- 0.52 e^{2 t})$) as
shown in Fig. 1(d), the amplitude and width becomes constant for
large $t$. This means that the stability of the soliton is
ensured. If we choose the trapping strength to oscillate between
confining and expulsive regime ($\Omega(t)^2=-0.5 \cos(2t)+ 0.125
\sin(2 t)^2$) as shown in Fig. 1(e), both the amplitude and width
of the soliton oscillate but no amplification occurs. When we
increase the frequency of oscillations of trapping frequency
($\Omega(t)^2=-2 \cos(8 t) + 0.125 \sin(8 t)^2$) as shown in Fig.
1(f), the amplitude and width oscillates but the oscillation is
very small. In other words, the frequency of oscillation of the
amplitude is proportional to frequency of oscillation of
time-dependent frequency. This is similar to the Feshbach
resonance \cite{saito} technique in which the amplitude of
oscillating soliton is nearly a constant while increasing the
frequency of interaction strength. From the Figs. 1(a-f), we
observe that the longevity of soliton can be controlled by fine
tuning the trapping potential strength. One can realize the above
phenomenon in experiments as well. For example, in the case of
$^7$Li, keeping BEC the attractive interaction strength $g = 0.25$
and $\omega_z = 2 \pi \times 710$Hz,  one can oscillate the radial
trapping frequency approximately between $2 \pi \times 80$Hz and
$2\pi \times 200$Hz. Invoking this ideas in the two-soliton
solution, one observes an interesting splitting and recombination
phenomenon.

\begin{figure}
\epsfig{file=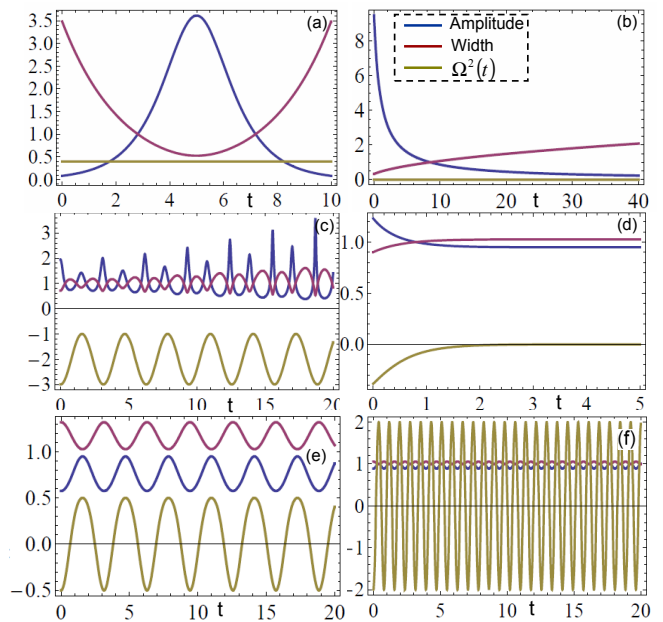, width=1.0\linewidth} \caption{(Color
online) The amplitude, width and trapping strength of one-soliton
solution plotted for different case. (a) For constant
time-independent trapping potential ($\Omega(t)^2=0.4$ (i.e., $
\omega_r=2\pi \times 284$Hz, $\omega_z=2\pi \times 710$Hz)) with
$g=0.0005, a=10+0.2i, b=10-0.3i$. (b) The trapping potential is
switched off when $g=0.05, a=1.5+0.2i, b=1.5-0.3i$. (c)
Oscillating time-dependent trap within confining regime
($\Omega(t)^2=-2 -cos(2t)$), (d) exponentially decreasing
confining trap approaching zero, (e) trapping frequency oscillates
between confining and expulsive regime and (f) increasing
oscillation frequency of the trap with $g = 0.05$, $a=0.15+0.2i,
b=0.15-0.3i$.}
\end{figure}

\section{Split and Recombincation of two-soliton}

To generate two-soliton solution, we consider $N$ = 2 in the
expression $\bar{g}^{(1)}= \sum_{j=1}^{N} e^{\chi_j}$ in Eq. (10)
and repeat the same procedure as in one soliton case, to get
$\bar{g}^{(1)}$, $\bar{g}^{(3)}$, $f^{(2)}$ and $f^{(4)}$ while
the series gets truncated at $\bar{g}^{(3)}$, $f^{(4)}$. The
two-soliton solution can be explicitly written as
\begin{equation}
\psi_2 = e^{i \alpha (t) \frac{(x^2 +y^2)}{2}} \frac{\bar{g}^{(1)}+\bar{g}%
^{(3)}}{1+f^{(2)}+f^{(4)}},
\end{equation}
where
\begin{eqnarray}
\bar{g}^{(1)} &=& e^{\chi_1} + e^{\chi_2}, \\
\bar{g}^{(3)} &=& e^{\chi_1+ \chi_1^* + \chi_2 +
\Gamma_1}+e^{\chi_1 +
\chi_2 +\chi_2^*+\Gamma_2}, \\
f^{(2)} &=& e^{\chi_1 + \chi_1^* + \eta_1}+e^{\chi_2 + \chi_2^*
+\eta_2}\nonumber\\
&+&e^{\chi_1 +\chi_2^* + \eta_0}+e^{\chi_1^* + \chi_2 +\eta_0^*},
\\
f^{(4)}&=&m(t) e^{\chi_1 + \chi_1^* + \chi_2 + \chi_2^*},
\end{eqnarray}
and
\begin{eqnarray}
\chi_i &=& h_1^{(i)}(t) x +h_2^{(i)}(t) y + h_3^{(i)}(t); \quad i =1,2, \\
h_1^{(1)}(t) &=& a e^{-\int \alpha(t) dt}; \quad h_2^{(1)} = b
e^{-\int
\alpha(t) dt}, \\
h_1^{(2)}(t) &=& c e^{-\int \alpha(t) dt}; \quad h_2^{(2)} = d
e^{-\int
\alpha(t) dt}, \\
h_3^{(1)}(t)&=&\int(\frac{i}{2} (a^2+b^2)
e^{-2\int\alpha(t)dt}-\alpha(t))dt,
\\
h_3^{(2)}(t)&=&\int(\frac{i}{2} (c^2+d^2)
e^{-2\int\alpha(t)dt}-\alpha(t))dt,
\end{eqnarray}
\begin{eqnarray}
e^{\Gamma_1}&=&\delta_1 e^{\eta_1}+\delta_2 e^{\eta_0^*}, \quad
e^{\Gamma_2}=\delta_3 e^{\eta_0}+\delta_4 e^{\eta_2}, \\
e^{\eta_1}&=&ge^{2\int \alpha(t) dt}/((a+a^*)^2+(b+b^*)^2), \\
e^{\eta_2}&=&ge^{2\int \alpha(t) dt}/((c+c^*)^2+(d+d^*)^2), \\
e^{\eta_0}&=&ge^{2\int \alpha(t) dt}/((a+c^*)^2+(b+d^*)^2), \\
e^{\eta_0^*}&=&ge^{2\int \alpha(t) dt}/((c+a^*)^2+(d+b^*)^2),
\end{eqnarray}
\begin{eqnarray}
\delta_1 &=&\frac{(c - a) (a + a^*) + (d - b) (b + b^*)}{(c + a^*)
(a + a^*)
+ (d + b^*) (b + b^*)}, \\
\delta_2 &=&\frac{(a - c) (c + a^*) + (b - d) (d + b^*)}{(c + a^*)
(a + a^*)
+ (d + b^*) (b + b^*)}, \\
\delta_3 &=&\frac{(c - a) (a + c^*) + (d - b) (b + d^*)}{(c + c^*)
(a + c^*)
+ (d + d^*) (b + d^*)}, \\
\delta_4 &=&\frac{(a - c) (c + c^*) + (b -d) (d + d^*)}{(c + c^*)
(a + c^*)
+ (d + d^*) (b + d^*)}, \\
m(t)&=&\frac{1}{4} \frac{e^{4\int \alpha(t) dt}(a -c)^2 (a^* -
c^*)^2}{(a + a^*)^2 (c + a^*)^2
(a + c^*)^2 (c + c^*)^2)}.\nonumber\\
\end{eqnarray}
The constraint $a=b$ and $c=d$ ensures the two-soliton solution to
be exact and the solitons are parallel to each other so called
``parallel bright matter-wave solitons".

\begin{figure}
\epsfig{file=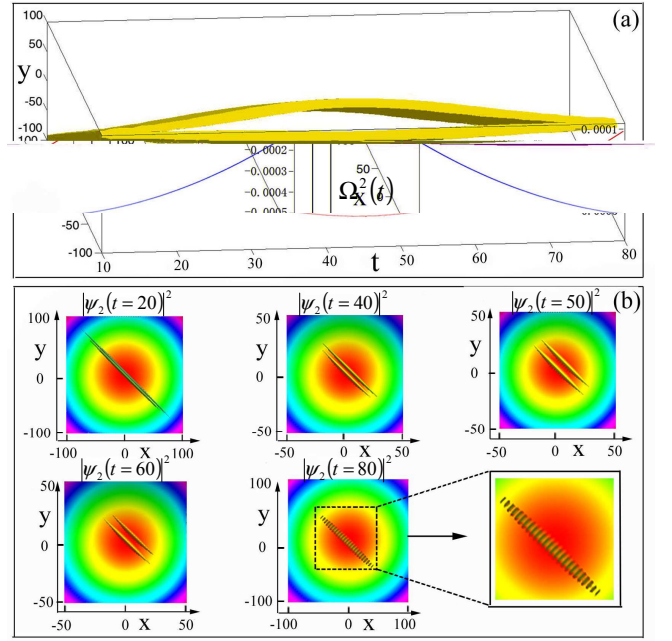, width=1.0\linewidth} \caption{(Color
online) (a) The split-recombination of two parallel bright
matter-wave solitons with the following parameters $a =b=0.3+0.1
i, c=d=0.3, \alpha(t) = -0.001 (t-50)$. (b) The snapshots of
two-soliton along $z$-axis at different instants of time. The trap
strength $\Omega(t)^2$ parabolically varying between $-0.0001$ and
$-0.0005$ in a span of 80 ms.}
\end{figure}

From the two-soliton solution (Eq. (17)), one can observe the
dynamics of parallel bright matter-wave solitons in the
two-dimensional harmonic trap. We choose the function
$\alpha(t)=-0.001(t-50)$ in which the trap strength
($2\Omega(t)^2=\alpha'(t)+\alpha(t)^2$) varies within the
confining regime ($\Omega(t)^2<0$) between $t\simeq10$ and
$t\simeq80$. First, we consider a two parallel bright solitons
merged in a single bound state with the initial conditions
$a=b=0.3+0.1 i, c=d=0.3,$ and attractive interaction strength is
kept constant at $g=0.5$. We then modulate the trap strength in a
parabolic manner as shown in Fig. 2(a). From Fig. 2(a), we observe
that a soliton coherent state at $t = 10$ evolves in time and
splits into two parallel solitons. Later on, they recombine at $t
= 80$ and produce an interference pattern as shown in Fig. 2(b).
The formation of interference pattern due to the collision of two
parallel bright solitons with zero angle of incidence in a single
component condensate is different from the interference pattern
produced by the expansion of two condensate clouds \cite{andre}
when the trapping potential is switched off. Thus, our results
reinforce the fact that the matter-waves originating from the
condensates (bright solitons) do interfere and produce a fringe
pattern as a clear signature of the long range spatial coherence
of the condensates. This is another way of generating interference
pattern and one can delay or control the creation of pattern by
properly modulating the trapping potential strength. The two
parallel bright soliton maximum separation depends on the strength
of trapping potential as shown in Fig. 2(a). Experimentally, one
can realize this phenomenon by parabolically modulating the
trapping frequency within the two-dimensional confining regime.

In experiment, to generate parallel bright solitons in a
two-dimensional harmonic trapping potential, initially the
trapping frequencies in the case of $^7$Li BECs are $\omega_r = 2
\pi \times 50$Hz and $\omega_z = 2 \pi \times 710$Hz with the
effective attractive interaction being  $g = 0.25$. This trap can
be determined by a combination of spectroscopic observations,
direct magnetic field measurement and the observed spatial
cylindrical symmetry of the trapped atom cloud \cite{rycht}. After
making this set up to generate two-soliton, the condensates split
into two parts using radio frequency (RF)-dressed potential
\cite{schum}. This technique has the advantage of allowing a
smooth transition from a single trap into a double well potential
and hence coherent splitting is possible with splitting range from
$3-80\rm \mu m$. Due to the constant attractive scattering length,
the two-bright solitons are generated from two nearly separated
condensates. We may consider this as a initial coherent state of
two-soliton split-recombination phenomenon. Then, one starts to
tune the radial trap frequency in a parabolic manner after
switching off the RF. When the radial trapping frequency which
after fine tuning reaches the initial value, the two-bright
solitons recombine and produce fringe pattern.

\section{Conclusion}

We have found exact bright matter-wave solitons for a
two-dimensional BEC with attractive interaction strength in a
time-dependent harmonic trap. Even though attractive interaction
strengths usually lead to the collapse of the condensates, one can
sustain the stability of solitons by properly modulating the
frequency of the radial trap. Our method is different from
stabilizing solitons by Feshbach resonance technique. We also
observe that the splitting and recombination of matter-wave
solitons which generates interference pattern in the process is a
new phenomenon in two dimensional BECs and may have wider
ramifications in the manipulation of atoms in BECs. In particular,
ultra-narrow two-dimensional BEC solitons is very useful in the
field of nanolithography.

\section*{Acknowledgements} This work was supported by the
International Young Scientist Fellowship of IOP, CAS under the
Grant No. 2010002 and NSFC under grants Nos. 10874235, 10934010,
60978019, the NKBRSFC under grants Nos. 2009CB930701, 2010CB922904
and 2011CB921500. RR acknowledges DST and DAE-NBHM for financial
support.

\end{document}